\documentclass[%
preprint,
superscriptaddress,
nofootinbib,
aps,
longbibliography,
]{revtex4-1}

\usepackage{CJKutf8}

\usepackage{graphicx}
\usepackage{siunitx}
\usepackage{stmaryrd}	
\usepackage{amssymb,amsmath,amsfonts,latexsym,mathrsfs,amsthm}
\usepackage{natbib}
\usepackage{subfigure}

\newcommand{\cref}[0]{C_{\rm{ref}}}

\newcommand{\qtot}[0]{Q_{\rm{tot}}}

\newcommand{\subl}[1]{{#1}_{\rm L}}

\newcommand{\subr}[1]{{#1}_{\rm ref}}
\newcommand{\subh}[1]{{#1}_{\rm H}}

\newcommand{\subc}[1]{{#1}_{\rm c}}
\newcommand{\subz}[1]{{#1}_{\rm 0}}

\newcommand{\jump}[1]{\left\llbracket{#1}\right\rrbracket}

\begin{document}
	
	
	\title{Impact of Leakage for Electricity Generation by  Pyroelectric Converter}
	\begin{CJK*}{UTF8}{gbsn}
		\author{Chenbo Zhang (张晨波)}
		\affiliation{Department of Mechanical and Aerospace Engineering, Hong Kong University of Science and Technology, Clear Water Bay, Hong Kong}
		\affiliation{HKUST Jockey Club Institute for Advanced Study, Hong Kong University of Science and Technology, Clear Water Bay, Hong Kong}
		\author{Zhuohui Zeng (曾卓晖)}
		\affiliation{Department of Mechanical and Aerospace Engineering, Hong Kong University of Science and Technology, Clear Water Bay, Hong Kong}
		\author{Zeyuan Zhu (朱泽远)}
		\affiliation{Department of Mechanical and Aerospace Engineering, Hong Kong University of Science and Technology, Clear Water Bay, Hong Kong}
		\author{Mostafa Karami }
		\affiliation{Department of Mechanical and Aerospace Engineering, Hong Kong University of Science and Technology, Clear Water Bay, Hong Kong}
		\author{Xian Chen (陈弦)}
		\email{xianchen@ust.hk}
		\affiliation{Department of Mechanical and Aerospace Engineering, Hong Kong University of Science and Technology, Clear Water Bay, Hong Kong}
		
		\date{\today}
		
		\begin{abstract}
			Pyroelectric energy converter is a functional capacitor using pyroelectric material as the dielectric layer. Utilizing the first-order phase transformation of the material, the pyroelectric device can generate adequate electricity within small temperature fluctuations. However, most pyroelectric capacitors are leaking during energy conversion. In this paper, 
			we analyze the thermodynamics of pyroelectric energy conversion with consideration of the electric leakage. Our thermodynamic model is verified by experiments using three phase-transforming ferroelectric materials with different pyroelectric properties and leakage behaviors. We demonstrate that the impact of leakage for electric generation is prominent, and sometimes may be confused with the actual power generation by pyroelectricity. We discover an ideal material candidate, (Ba,Ca)(Ti,Zr,Ce)O$_3$, which exhibits large pyroelectric current and extremely low leakage current. The pyroelectric converter made of this material generates 1.95$\mu$A/cm$^2$ pyroelectric current density and 0.2 J/cm$^3$ pyroelectric work density even after 1389 thermodynamic conversion cycles.

		\end{abstract}
		
		
		\maketitle
	\end{CJK*}
	
	\section{Introduction}
	
Energy harvesting from waste heat is an emerging research field in sustainable energy science \cite{garcia2018review}. The heat, particularly in low-grade regime (\emph{i. e.} $< 250^\circ$C) is abundant in nature from solar radiation and geothermal activities. Industries and human daily activities generate an immense amount of waste heat below $200 ^\circ$C. Ferroelectric materials with the temperature-dependent polarization attract increasing attention in recent years for electricity generation from 40$^\circ$C to 150$^\circ$C \cite{Olsen1985, sebald2009, yang2012pyro, Pandya2018}. The mechanism of this energy conversion is commonly characterized as the pyroelectric effect. Many reported energy converters are made of a capacitor with a dielectric layer whose pyroelectric property is designed for harvesting low-grade heat. Most of them are thin-film devices \cite{yang2012pyro, Pandya2018} working under a bias electric field, provided by an external power. The electric energy generated by these devices is usually confused with the applied bias electric field. Without such a bias electric field, the performance of the energy conversion becomes rather marginal \cite{Olsen1985, yang2012pyro}.
	
When utilizing the first-order phase transformation to pyroelectric energy conversion, the converted electricity can be drastically magnified due to the abrupt jump of polarization caused by the structural change from ferroelectric to paraelectric phases. Without any applied bias field, the generated pyroelectric current by a bulk phase-transforming capacitor, \emph{i.e.} 150 nA \cite{Zhang2019prapplied} is larger than the current generated by the state-of-art thin film pyroelectric converter, \emph{i.e.} 40 nA \cite{Pandya2018} almost by a factor of four. Similar energy conversion device was demonstrated using a single-crystal Barium Titanate \cite{bucsek2019direct}, which generates 400 nA pyroelectric current through only a 5K temperature difference covering its tetragonal (ferroelectric) to cubic (paraelectric) phase transformation. This has been by-far the largest reported pyroelectric current purely converted from heat. 
	
Removal of the bias electric field re-defines the thermodynamic cycle for pyroelectric energy conversion from small temperature fluctuations. Depending on the initial condition,  a couple of new thermodynamic models were proposed to analyze the power output for the power-source-free pyroelectric generators \cite{Zhang2019prapplied, bucsek2019direct, bucsek2020}. In our previous study for phase-transforming pyroelectric energy conversion, the initial state of the thermodynamic system is set as $(\subl P, \subl E, \subl T)$ for temperature $\subl T$ lower than transition temperature, the working conjugate of polarization $P_L$ and electric field $\subl E > 0$ at ferroelectric phase. The initial polarization is achieved by an external DC voltage source. For the bulk capacitor, the applied DC voltage is in the range of 30V - 50V \cite{Zhang2019prapplied, bucsek2019direct}. When the ferroelectric capacitor is balanced with its charge reservoir, the external DC voltage source is removed and the system is assumed to be electrically isolated. When the ferroelectric material undergoes a phase transformation, the charges start flowing. Therefore, this is a constant-current-output energy conversion device. Ideally, the total charges in the isolated system preserve if the phase-transforming dielectric layer is a perfect insulator. The charges leakage is unavoidable and non-negligible especially at elevated temperature for the perovskite oxides \cite{bucsek2020, nagaraj1999leakage, pabst2007leakage, pintilie2007ferroelectric, yang2007temperature, sun2017leakage}. Over numerous energy conversion cycles, the performance of the device degrades significantly, which highly hinders the real applications of this new energy conversion technology.  For the thin-film capacitors, the leakage current is a more critical problem for electricity generation \cite{pabst2007leakage,sun2017leakage, ke2010oxygen}. 
	
\begin{figure*}[ht]
\includegraphics[width = 0.48\textwidth]{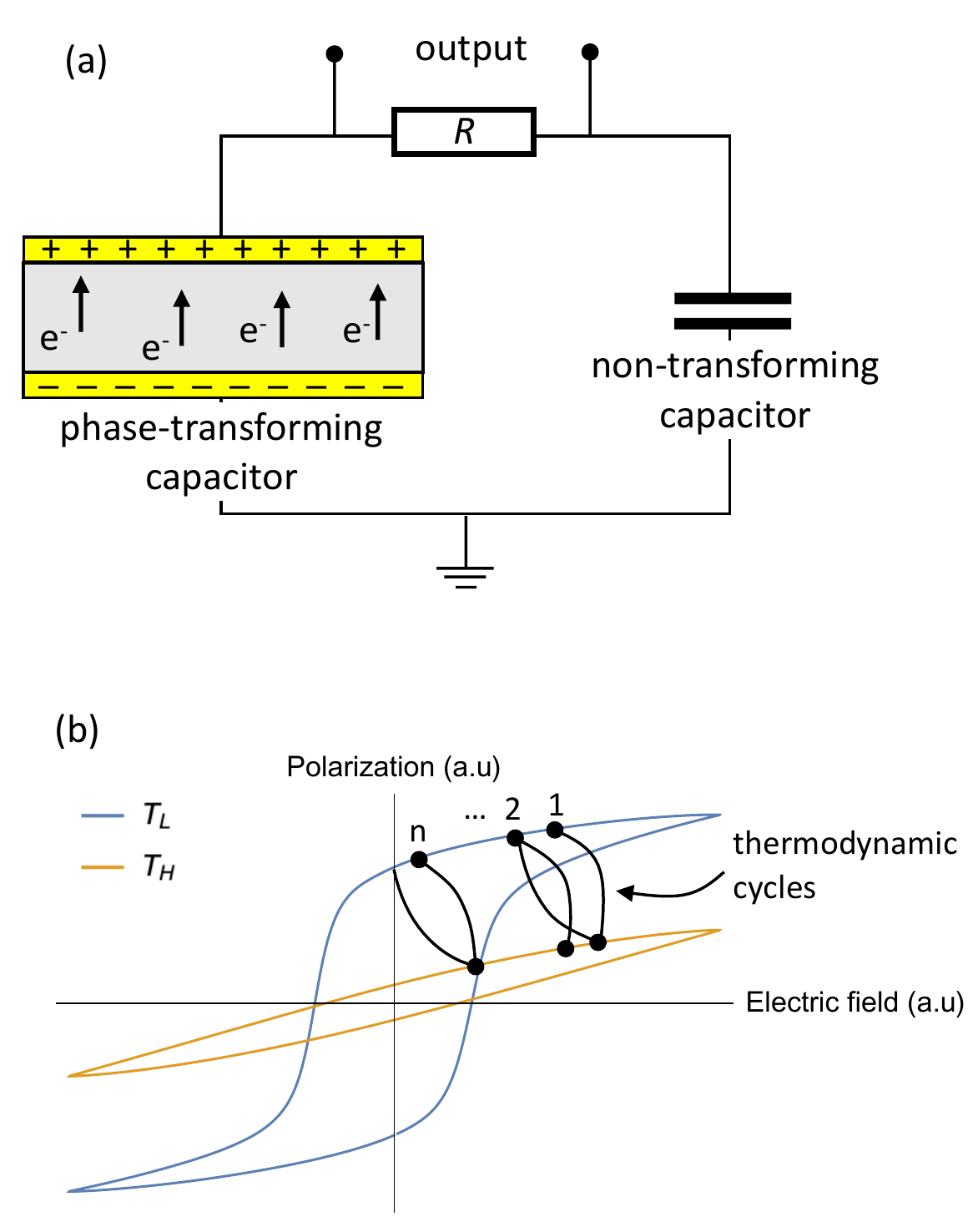}
\caption{(a) Schematic of the energy conversion system using the phase-transforming capacitor; (b) Schematic of polarization versus electric field at ferroelectric ($\subl T$) and paraelectric ($\subh T$) phases by which the power-source-free thermodynamic cycles \cite{Zhang2019prapplied} is designed with consideration of the leakage-induced degradation.}\label{fig:schematic}
\end{figure*}

This paper theorizes the impact of leakage to the power-source-free energy conversion system shown in Fig.~\ref{fig:schematic}(a). The charges leak inside the dielectric layer as the energy conversion begins. Unlike the ideal insulation system, the electric field of the phase-transforming planar capacitor can not reach its initial value after each of the energy conversion cycles, as illustrated in Fig. \ref{fig:schematic}(b). 

\section{Thermodynamic analysis}

We consider the phase-transforming capacitor as the thermodynamic system. When the system thermally fluctuates about the phase-transformation temperature $\subc T$, the charges $Q$ held by the capacitor accordingly fall and rise due to the alternation of capacitance caused by the first-order phase transformation between ferroelectric and paraelectric phases. The charge can be expressed as a function of the temperature of the phase-transforming capacitor, $T$, and the voltage on the capacitor, $V$, \emph{i.e.} $Q = Q(V, T)$. During the cyclic thermodynamic processes between the states $(\subl Q, \subl V, \subl T)$ and $(\subh Q, \subh V, \subh T)$ where $\subl T < \subc T < \subh T$, all three state variables $(Q, V, T)$ change simultaneously without an application of a bias electric field \cite{Zhang2019prapplied}.  Considering a planar capacitor, the voltage is proportional to the electric field in the capacitor. Let $t=0$ mark the beginning of the first thermodynamic cycle (cycle 1 in Fig. \ref{fig:schematic}(b)), the electric field in the capacitor is 
\[
E(0) = E_0 = \frac{V_0}{d},
\]
where $\subz V$ is gained by connecting the whole electric circuit in Fig.\ref{fig:schematic}(a) to an external DC source. Let $t>0$ mark the energy conversion processes after the removal of external DC source. Due to electric leakage, the stored charges $Q(V, T)$ can not reach the initial state of its previous thermodynamic cycle, as in Fig. \ref{fig:schematic} (b). In this case, the relative starting state of the next cycle drifts down along the polarization curve. As a result, the thermodynamic system is cycle-dependent. Let time $t_n$ label the starting state of the $n^\text{th}$ cycle, 
\[
E(t_{n+1}) < E(t_{n}) < E(t_{n-1}) < \ldots < E_0.
\]
Consequently, the stored charges in ferroelectric phase decrease after every cycle. If such a leakage continues, there would be no charges left after a finite number of energy conversion cycles, which ceases the electricity generation process.  

\subsection{Governing equations of energy conversion}
	
At $t=0$, the total charges in the system are calculated by 
\begin{equation} 
\qtot = \subr C \subz V + Q(\subz V,\subl T), \label{Qtot}
\end{equation} 	
where $\subr C$ denotes the capacitance of the non-transforming capacitor, considered as the reference capacitor in our experiment. At $t > 0$, the charge balance can be expressed as 
\begin{equation} 
\qtot-\int_{0}^{t} \subc I(\xi) \text{d}\xi =\subr Q(t)+Q(V(t),T(t)),\label{Qbalance}
\end{equation} 	
where $\subr Q$ is the time-dependent function of charges held by the reference capacitor for $t > 0$, the function $\subc I$ represents the charge leakage current in the phase-transforming ferroelectric capacitor, which is non-zero after energy conversion starts. 
	
Considering the electric circuit in Fig. \ref{fig:schematic}(a), we have
\begin{equation} 
V(t) = V_\text{d}(t) + \frac{\subr Q(t)}{\subr C} \label{Vbalance},
\end{equation} 
where $V_\text{d}$ is voltage on the load resistor.  
Substituting \eqref{Vbalance} into \eqref{Qbalance} and taking the first order time derivative, we get
\begin{equation} 
\dot{Q}+\subr C\dot{V}+\subc I-\dot{V}_\text{d}\subr C=0. \label{firstgovern}
\end{equation} 	
Since $Q$ is a function of $V$, the equation \eqref{firstgovern} is a first order ODE about variables $V$ and $V_\text{d}$. The voltage on the load resistor is attributed to the generation of current $-\dot{Q}$ and the leakage current $\subc I$, as
\begin{equation}
    V_\text{d} = -(\dot Q + \subc I) R.
\end{equation} 
Therefore, the governing equation can be written as 
\begin{equation} 
R\subr C\ddot{Q} + \dot{Q} + \subr C \dot{V} + \subc I + R \subr C \dot{\subc I} = 0. \label{secondgovern}
\end{equation} 
The equation \eqref{secondgovern} is a second order ODE about the variable $V$. We assume that the applied voltage across the thin plate phase-transforming pyroelectric capacitor is linearly related to the electric field $E$ by $V = E d$ where $d$ denotes the thickness of the thin plate. The constitutive relations for $Q$ and $\subc I$ can be also characterized as $Q(E,T) = A q(E,T)$ and $\subc I(E,T) = A J(E,T)$ where $A$ is the area of the planar capacitor, $q(E,T)$ and $J(E,T)$ are the charge density and leakage current density.
Modulated by the area $A$, the equation \eqref{secondgovern} becomes
\begin{equation} 
R\subr C\ddot{q} + \dot{q} + \frac{d}{A}\subr C \dot{E} + J + R \subr C \dot{J} = 0. \label{qgov}
\end{equation} 	

\subsection{Characterizations of ferroelectric properties}
According to the constitutive relation \cite{ferroelectricbook2017}, the charge density of phase-transforming pyroelectric capacitor is a function of electric field $E$ and temperature $T$,
\begin{equation}
    q = \subz \varepsilon E + P(E,T), \label{q}
\end{equation}
where $\subz \varepsilon$ is the vacuum permittivity and the constitutive function $P(E,T)$ is the polarization of the phase-transforming ferroelectric material. The first and second order time derivatives can be calculated directly from \eqref{q} with the partial differential quantities 
$P_{,E}=\frac{\partial P}{\partial E}$, $P_{,T}=\frac{\partial P}{\partial T}$, $P_{,EE}=\frac{\partial^2 P}{\partial E^2}$,  $P_{,TT}=\frac{\partial^2 P}{\partial T^2}$, and  $P_{,TE}=\frac{\partial^2 P}{\partial T \partial E}$, in which $\kappa=P_{,T}$ is an important ferroelectric property, defined as the pyroelectric coefficient. It has been widely used to judge the performance of the pyroelectric generators \cite{sebald2009, bowen2014pyroelectric, Pandya2018, pandya2018pyroelectric}. Note that, when a first-order phase transformation is coupled with pyroelectric property, these derivatives vary as the temperature and electric field.
    
\begin{figure}[ht]
\includegraphics[width=0.45\textwidth]{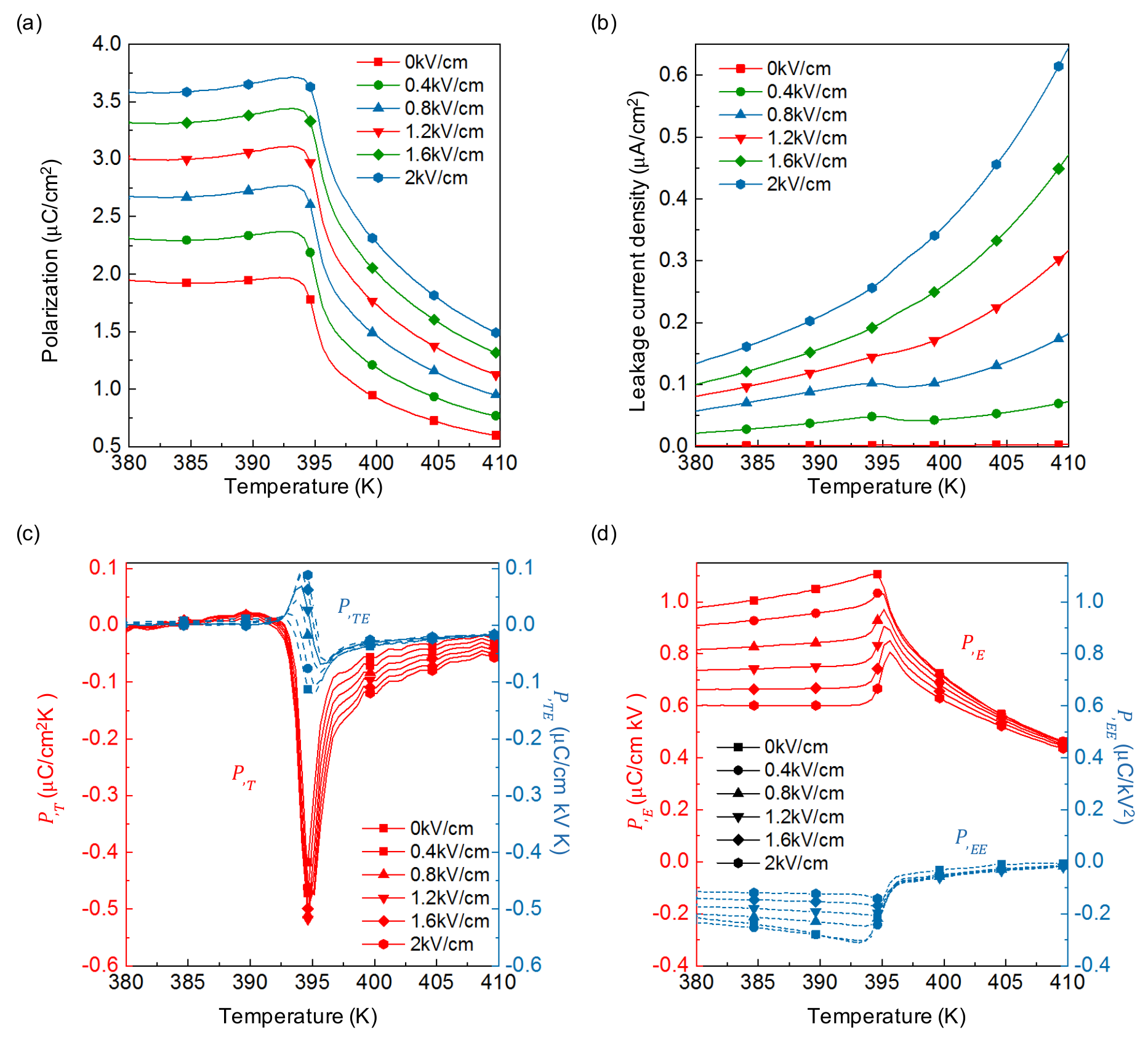}
\caption{The temperature-dependent (a) polarization and (b) leakage current density of the phase-transforming BaTiO$_3$ under electric fields from 0kV/cm to 2kV/cm in the temperature range from 380K to 410K,  corresponding to the derivatives (c) $P_{, T}$ (red), $P_{,TE}$ (blue) and (d) $P_{, E}$ (red), $P_{,EE}$ (blue). }
\label{Fig2_BTOPTLT}
\end{figure}

The constitutive relations $P(E,T)$ and $J(E,T)$ are experimentally characterized by the aixACCT TF ferroelectric analyzer 2000E system. Fig.~\ref{Fig2_BTOPTLT}(a) and (b) shows temperature dependent ferroelectric polarization and leakage current density in the polycrystal BaTiO$_3$(BTO) under different electric fields from 0kV/cm to 2kV/cm. The electric leakage of the material becomes more prominent at the elevated temperature under the larger applied electric field. In our energy conversion experiment, the working electric field on pyroelectric capacitor is between 0.5kV/cm and 1.2kV/cm. When the applied electric field is higher than 1.2kV/cm, the increasing of leakage current becomes much rapid at elevated temperature regime. This indicates that the leakage problem is worse for the pyroelectric thin-film converters working under biased fields \cite{yang2012pyro, Pandya2018, pandya2018pyroelectric}. 
In the characterized temperature regime, the polarization of the material decreases as temperature increasing with an abrupt drop at 395K, which reveals the morphotropic phase boundary between tetragonal and cubic. Evidently, Fig. \ref{Fig2_BTOPTLT}(c) shows the singularity of the pyroelectric coefficient (\emph{i. e.} $P_{, T}$) at 395K. Similarly, the derivative $P_{,E}$ is also singular at the phase-transforming temperature. Due to such a singularity, the conventional formula of pyroelectric current \cite{bowen2014pyroelectric} can not be applied in the vicinity of the morphotropic phase boundary when first-order phase transformation occurs. Instead, the Clausius-Clapeyron relation should be considered in the thermodynamic analysis of energy conversion \cite{Zhang2019prapplied, bucsek2019direct, bucsek2020}. 
The second order derivatives $P_{,EE}$ and $P_{,TE}$ are calculated numerically, as shown in Fig.~\ref{Fig2_BTOPTLT}(c) and (d), which are very small compared to the first derivative functions. For simplicity, we neglect these second order derivatives in the numerical solutions of the governing equation \eqref{qgov}. Note that we keep the second order $P_{,TT}$ in all computations. 

After dropping terms of $P_{,TE}$ and $P_{,EE}$, the governing equation \eqref{qgov} becomes
\begin{equation} 
(\subz\varepsilon+P_{,E})\ddot{E} + C_1\dot{E} + C_2 = 0, \label{govern}
\end{equation} 
where	
\begin{align} 
&C_1 = \frac{1}{R\subr C}(\subz\varepsilon +P_{,E}) + \frac{d}{A R} + J_{,E}, \label{con1} \\
&C_2 = P_{,T}\ddot{T}+P_{,TT}\dot{T}^2+(\frac{P_{,T}}{R\subr C}+J_{,T})\dot{T}+\frac{J}{R\subr C}. \label{con2}
\end{align} 
It can be solved numerically under the initial conditions
\begin{equation} 
E(0)=\subz E=\frac{\subz V}{d}, \ T(0)=\subl T,\ \dot{E}(0) = 0. \label{initial}
\end{equation} 

In order to benchmark materials of different geometry under various experimental setting, the dimensionless variables are used to represent the governing equation by the symbol with an overhead bar, $\bar{\cdot}$. For a specific phase-transforming pyroelectric planar capacitor with thickness $d$, the initial charging voltage $V(0) = V_0$ associated with the electric field $E_0 = V_0/d$ at temperature $\subl T$ lower than the phase transition temperature, the nondimensional voltage, electric field, temperature and time are defined as
\begin{equation} \label{eq:dim1}
\bar{V}=\frac{V}{\subz V},\ \bar{E}=\frac{E}{\subz E}, \ \bar{T}=\frac{T-\subl T}{\subh T-\subl T}, \ \bar{t}=\frac{t}{\tau},
\end{equation} 
where the working temperature range of the device is between $\subl T$ and $\subh T$ such that $\subl T < \subc T < \subh T$. It takes $\tau$ seconds per cycle, which fully covers the duration of phase transformation. According to \eqref{eq:dim1}, the values of $\bar V$, $\bar E$, $\bar T$ and $\bar t$ are all between 0 and 1.
Let $\subl P = P(\subz E, \subl T)$ be the initial polarization in ferroelectric phase, and $\subh J = J(\subz E, \subh T)$ be the highest leakage current density, the nondimensional polarization and leakage current density are defined as 
\begin{equation}\label{eq:Pdim}
    \bar{P}=\frac{P}{\subl P}, \bar{J}=\frac{J}{\subh J} \in [0, 1]. 
\end{equation}
The corresponding derivatives are
\begin{align} 
&\bar P_{, \bar T} = \frac{P_{,T}}{\subl P} (\subh T - \subl T), \nonumber\\
&\bar P_{, \bar E} = \frac{E_0}{\subl P} P_{, E}, \nonumber\\
&\bar P_{, \bar T\bar T} = \frac{P_{,TT}}{\subl P} (\subh T - \subl T)^2, 
\end{align} 
and
\begin{equation} 
\bar J_{, \bar E}=\frac{\subz E}{\subh J}J_{,E}, \ 
\bar J_{, \bar T}=\frac{\subh T-\subl T}{\subh J}J_{,T}.\label{eq:Jdim}
\end{equation} 

Substitute \eqref{eq:dim1}, \eqref{eq:Pdim} and \eqref{eq:Jdim} into (\ref{govern}-\ref{con2}), we get the nondimensionalized governing equation
\begin{equation} \label{eq:gov_n}
\frac{(\subz\varepsilon \subz E + \subl P \bar{P}_{,\bar{E}})}{\tau^2} \ddot{\bar{E}} + \frac{\bar C_1}{\tau}\dot{\bar{E}} + \bar{C_2} = 0,
\end{equation} 
where
\begin{align} 
   \bar C_1 =& \frac{\subz \varepsilon \subz E + \subl P \bar{P}_{,\bar{E}}}{R\subr C} + \frac{d E_0}{A R} + \subh J \bar J_{,\bar E}, \\
\bar C_2 =& \frac{\subl P \bar{P}_{,\bar T}\ddot{\bar T} + \subl P \bar P_{,\bar T \bar T} \dot{\bar T}^2}{\tau^2} +\nonumber\\
&(\frac{\subl P \bar P _{,\bar T}}{\tau R \subr C} + \frac{\subh J\bar J_{,\bar T}}{\tau})  \dot{\bar T} + \frac{\subh J \bar J}{R \subr C},
\end{align}
which can be solved numerically under the initial conditions
\begin{equation}
    \bar E(0) = 1, \ \bar T(0) = 0, \ \dot{\bar E}(0) = 0.
\end{equation}

\section{Experiment}

\begin{figure}[ht]
\includegraphics[width=0.45\textwidth]{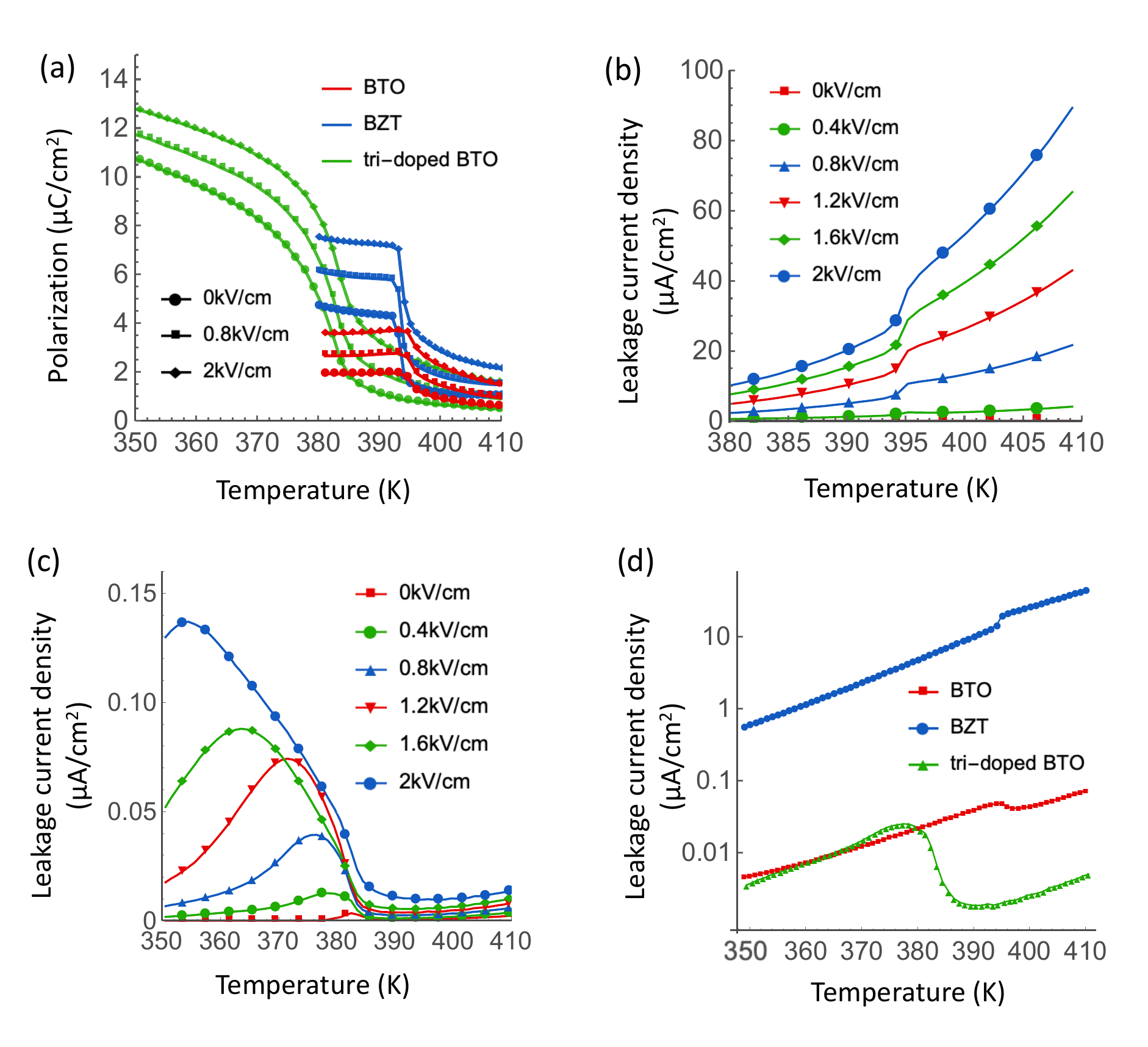}
\caption{(a) The temperature-dependent polarization of BTO (red), BZT (blue) and tri-doped BTO (green), and the temperature-dependent leakage current density in (b) BZT and (c) tri-doped BTO under electric fields from 0kV to 2kV/cm. (d) Comparison of the leakage current density among three capacitors made of BTO (red), BZT (blue) and tri-doped BTO (green) under 30V voltage.}
\label{Fig3_PTLT}
\end{figure}

We use three Barium Titanate based ferroelectric materials to investigate the influences of electric leakage on pyroelectric energy conversion, and to verify our numerical model in \eqref{eq:gov_n}. They are polycrystal BaTiO$_3$(BTO), Ba(Ti, Zr$_{0.01}$)O$_3$ (BZT) and (Ba, Ca$_{0.05}$)(Ti, Zr$_{0.01}$, Ce$_{0.01}$)O$_3$ (tri-doped BTO). The BTO and BZT are synthesized by the conventional ceramic sintering method with the detailed procedure introduced in reference \cite{Zhang2019prapplied}, whose ferroelectric property was characterized in Fig. \ref{Fig3_PTLT}(a). It shows that the Zr addition increases the polarization in ferroelectric phase. As shown in Figs. \ref{Fig2_BTOPTLT} (b) and \ref{Fig3_PTLT} (b), both exhibit similar trend of leakage as temperature increasing under different electric fields. From energy conversion point of view, the figure of merit of BZT is higher than that of BTO. But the leakage of BTO is much smaller than BZT. The tri-doped BTO is synthesized by the floating-zone method, which is commonly used for growing single crystals or textured polycrystals \cite{kojima1995crystal, brown1964floating}. The Cerium addition to BTO may increase the grain size and change the photorefractive properties at high temperature \cite{dou1996Ce,yasmm2011ce}, which facilitates the growth of crystals with large grain size. In this study, we aimed at synthesizing a new pyroelectric material with higher energy conversion ability but lower electric leakage. 

The powders of CaCO$_3$(Alfa Aesar, 99.5\%), CeO$_2$(Alfa Aesar, 99.9\%), ZrO$_2$(Sigma Aldrich, 99\%), BaCO$_3$(Alfa Aesar, 99.8\%) and TiO$_2$(Alfa Aesar, 99.8\%) were weighed according to the stoichiometric formulation Ba$_{0.95}$Ca$_{0.05}$Ti$_{0.98}$Zr$_{0.01}$Ce$_{0.01}$O$_3$, dissolved in the ethanol and milled by zirconia balls in the planetary ball miller at 630rpm for 24hrs. The ball-milled solution was dried and calcined at 1000$^\circ$C for 10 hours to obtain the tri-doped Barium Titanate powder through solid state reaction. Under 30MPa hydrostatic pressure for 30 minutes, the powder was hot-pressed into a rod-shaped green body of 6mm diameter and 10mm length.
The green body was sintered to form the feed and seed rods in the four-mirror Infra-red furnace (Quantum Design IRF11-001-00) by passing through the focal point slowly with the axial speed 3mm/hr and the angular speed 3rpm. The feed and seed rods were aligned vertically for the floating zone experiment using the same Infra-red furnace at the growth speed 10mm/hr and relative angular speed 50rpm in air. The molten zone was created and monitored throughout the entire growth period. Finally, the as-growth rod was solidified and sliced along the growth direction.  

\section{Result}

Fig.~\ref{Fig3_PTLT}(a) compares the temperature-dependent polarization among BTO (red), BZT (blue) and tri-doped BTO (green) under different electric fields from 0kV/cm to 2kV/cm. Within the testing temperature range, all samples undergo the first-order phase transformation accompanying with an abrupt polarization jump, $\jump P$ from ferroelectric to paraelectric phases. Among them, the tri-doped BTO exhibits the largest polarization ($11\mu$C/cm$^2$) in ferroelectric phase, making it an ideal candidate for pyroelectric energy conversion.  In Fig. \ref{Fig3_PTLT}(c), the electric leakage of tri-doped BTO has been reduced magnificently, which is about 1/6 of value for BTO and 1/200 of value for BZT in ferroelectric phase. Under the working condition for energy conversion, the benchmark of leakage current density is shown in Fig. \ref{Fig3_PTLT}(d). In sharp contrast to BZT and BTO, the leaking in the new ferroelectric material is much suppressed even in the temperature regime higher than its phase-transition temperature. In addition, we observed that the morphotropic phase boundary of tri-doped BTO gets relatively more gradual than that of BZT, which indicates that the pyroelectric coefficient at phase transformation is moderate compared to BZT. The measured pyroelectric coefficients are 0.42$\mu$C/cm$^2$K for BTO, 1.89$\mu$C/cm$^2$K for BZT, and 0.87$\mu$C/cm$^2$K for tri-dope BTO. 

We used these three materials as the dielectric layer to fabricate planar capacitors with geometries 37.44mm$^2 \times 0.74$mm for BTO, 59.44mm$^2 \times 0.257$mm for BZT, and 66.59mm$^2 \times 0.445$mm for tri-dope BTO. Using the energy conversion electric circuit in Fig.~\ref{fig:schematic}(a), each of the phase-transforming capacitors is connected to a 100k$\Omega$ load resistor and a 50$\mu$F reference capacitor in series. Before energy conversion, we used an external DC voltage source (Tektronix PS280) to charge the entire system so that both phase-transforming capacitor and reference capacitor gain 30V voltage, and no electric charges flow in the circuit. Then, the external DC voltage source was disconnected at temperature $\subl T < \subc T$.  
 
  \begin{figure*}
 	\includegraphics[width=0.7\textwidth]{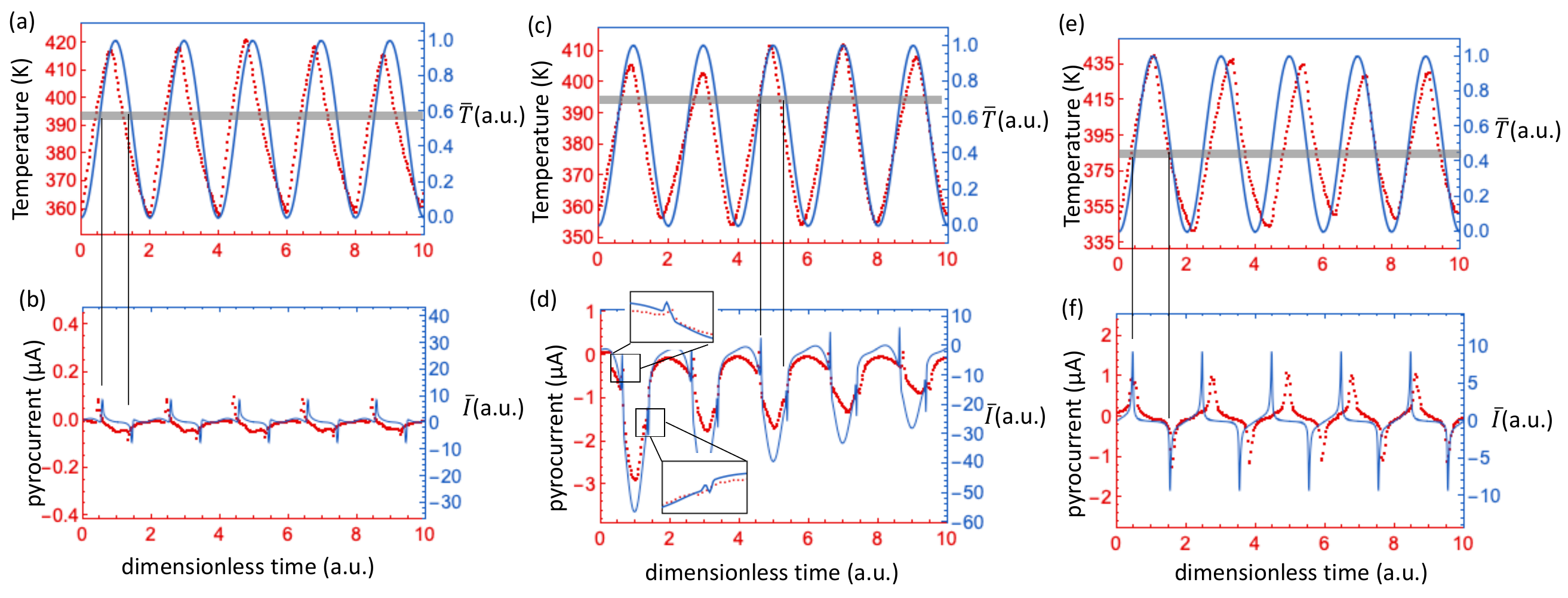}
 	\caption{Measured and simulated pyroelectric energy conversion by the temperature fluctuations in (a) BTO, (c) BZT, and (e) tri-dope BTO, corresponding to the pyrocurrent of (b)BTO, (d) BZT, and (f) tri-doped BTO. The experimental results are represented in red, while the simulated results are plotted in blue. }
 	\label{Fig4_expt}
 \end{figure*}

We used our custom-made thermal stage to drive temperature of the phase-transforming capacitor to achieve a periodic pattern around its transition temperature. For each of the phase-transforming capacitors, we collected five consecutive thermal cycles shown in Figs. \ref{Fig4_expt} (a), (c) and (e) corresponding to $\subl T = 360$K and $\subh T = 410$K. The total duration of all thermal cycles is 142 seconds for BTO, 115 seconds for BZT, and 63 seconds for tri-dope BTO, which are used to derive the nondimensionalized time variable. The normalized temperature patterns for these materials are plotted as blue curves in Figs. \ref{Fig4_expt} (a), (c) and (e). 

The pyrocurrent was characterized as the current passing through the load resistor, $R$, calculated as 
\begin{equation}
    I_p = i_p A,
\end{equation}
where 
\begin{equation}
    i_p = - (\dot{q} + J) = -\left[(\epsilon_0 + P_{,E})\dot{E} + P_{,T} \dot{T} + J\right]
\end{equation}
is the pyrocurrent density for an area $A$. 
It is attributed to the charge flow driven by the change of polarization and the internal leakage in the pyroelectric capacitor. As a result, the electric field across the pyroelectric capacitor is continuously decreasing throughout the energy conversion process. 
In experiment, the measured pyrocurrent (red) is shown in Fig. \ref{Fig4_expt} (b) BTO, (d) BZT and (f) tri-doped BTO, corresponding to the numerical solutions (blue) given by equation \eqref{eq:gov_n}. The generation of pyrocurrent in consecutive thermal cycles is well captured by our thermodynamic model for three samples with different ferroelectric properties. We conjecture that three factors play an important role in pyroelectric energy conversion by first-order phase transformation: 1) polarization jump before and after phase transformation; 2) pyroelectric coefficient at transition temperature; 3) leakage current density within the range of thermal fluctuations applied to the sample. For the BTO sample, all three factors are low, which indicates that it is a good insulator but a bad pyroelectric conversion material, corresponding to a very small pyrocurrent density (66.8 nA/cm$^2$) in all five cycles. The BZT sample is a good energy conversion material having the largest pyroelectric coefficient among three samples, and enhanced polarization in ferroelectric phase. But it is a bad insulator with non-negligible leakage. Considering a full loop of phase transformation, its averaged pyrocurrent density of the first cycle is 1.17$\mu$A/cm$^2$, which degrades to $0.34\mu$A/cm$^2$ after five cycles. Since the charge leakage is unidirectional, both measured and computed pyrocurrent for BZT are biased to the negative direction with respect to neutral point. The contribution of the leakage current to the work done on the load resistor sometimes covers the true electric energy converted by the pyroelectric material \cite{smith2017pyroelectric,you2018self,mistewicz2019sbsei}. For the tri-doped BTO, both pyroelectric capability and insulation are enhanced magnificently. As a result it shows 1.95$\mu$A/cm$^2$ pyrocurrent density without degradation at the end of fifth cycle. Note that the pyrocurrent achieved in tri-doped BTO is almost 3 times bigger than the largest pyrocurrent reported by far in single crystal Barium Titanate with polar axis aligned with the electric field \cite{bucsek2019direct}.

\section{Discussion}

\begin{figure}[ht]
	\includegraphics[width=0.45\textwidth]{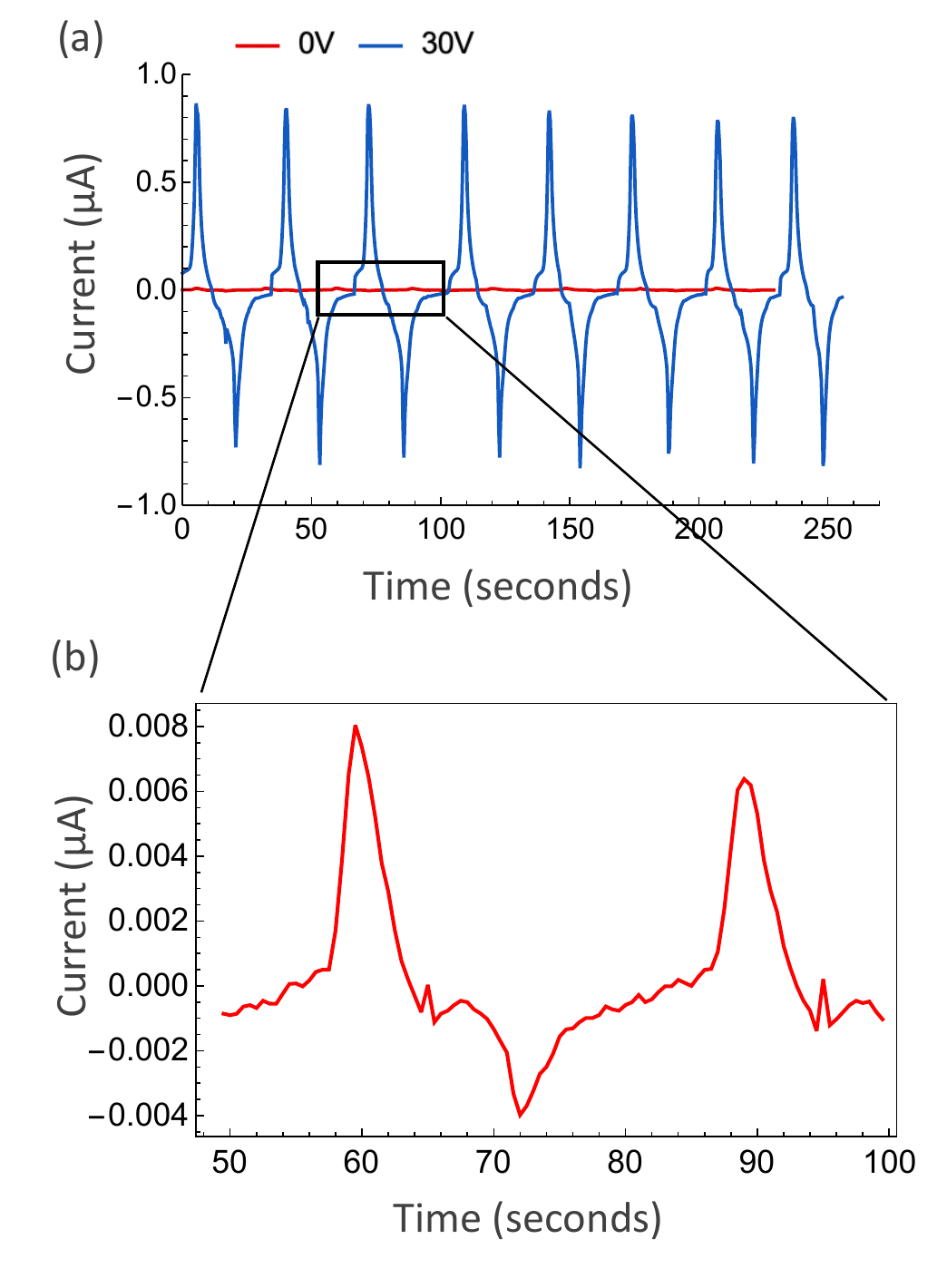}
	\caption{(a) Measured current generated by tri-doped BTO under zero electric field (red) and positive electric field $E > 0$ corresponding to 30V capacitor voltage. (b) Zoom-in region of the measured current under zero electric field.}
	\label{fig:nocharge}
\end{figure}

We use the least leaking pyroelectric capacitor (\emph{i.e.} tri-doped BTO) to conduct more cycles under different working conditions. Fig. \ref{fig:nocharge} compares the pyrocurrent given by the tri-doped BTO capacitor in a fully discharged state and a 30V charged state. Under the same experiment setup for energy conversion, the generated pyrocurrent strongly depends on the initial state $\subl Q(\subl V, \subl T)$. A fully discharged pyroelectric capacitor with $\subl V = 0$, produces much smaller electricity than that produced by the same capacitor with $\subl V = 30V$.
If the pyroelectric material is leaking, the capacitor keeps discharging while converting energy cyclically.  This intrinsic discharging eventually ceases the production of electricity in the system, even though the material itself is pyroelectrically functional. 

This opens a discussion for the effective life of a pyroelectric energy converter with intrinsic leakage. Let $\tau$ denote the duration of the half thermodynamic cycle from $(\subl P, \subl E, \subl T)$ to $(\subh P, \subh E, \subh T)$ where the temperature range $[\subl T, \subh T]$ fully covers the first-order phase transformation from ferroelectric to paraelectric. We assumed that the cooling process from $\subh T$ is timely symmetric to the first half thermodynamic cycle, then the electric field at the $N$th thermodynamic cycle yields
\begin{equation}
    E(2N\tau) < \ldots < E(2\tau) < E(0) = \frac{V_0}{d}.
\end{equation}
Solving the equation \eqref{eq:gov_n}, we can calculate the electric field $E(2N\tau)$ for $N \in \mathbb Z^+$. The effective life is defined as the number of thermodynamic cycles $N^\ast$, such that 
\begin{equation}
    E(2(N^\ast + 1) \tau) < \frac{V_0}{2 d} < E(2 N^\ast \tau).
\end{equation}
It means that after $N^\ast$ cycles, the electric field on pyroelectric capacitor has been reduced to half of its initially charged value. If we only consider the current output from the pyroelectric energy conversion, the output electric energy density on the load resistor is calculated as
\begin{equation}
    W(N) = \frac{AR}{d} \int_{0}^{2N\tau} \dot{q}^2 \text{d} t,
\end{equation}
where $\dot{q}$ denotes the current density passing through the load resistor $R$. Consider the work done by the electrostatic force in the pyroelectric capacitor during the cyclic thermodynamic processes, we define the pyro-work as   
\begin{equation}
    W_{\pi}(N) = - \int_{0}^{2 N \tau}  E \dot{P} \text{d}t. 
\end{equation}
Here the electric field should satisfy the governing equation \eqref{eq:gov_n} under the driving temperature profile $T(t)$. The polarization is evaluated by the experimentally characterized constitutive relations (Fig. \ref{Fig2_BTOPTLT} and \ref{Fig3_PTLT}). We calculate the effective life, the output work density and the pyro-work density for BTO, BZT and tri-doped BTO capacitors, summarized in Table \ref{tab:work}. 

Without leakage, the output electric energy is supposed to be the same as the electrostatic energy produced by the pyroelectric capacitor \cite{Zhang2019prapplied}. The loss of energy captured by our thermodynamic model is due to the internal loss of charges over cyclic energy conversion processes. For the most leaking capacitor BZT, its pyroelectric energy conversion ceases at the end of the 9$^\text{th}$ cycle. Until re-charging by an external battery, it won't produce electricity by pyroelectric effect even the pyroelectric material is still functional. Overall, BZT capacitor outputs $8.92 \times 10^{-4}$ J/cm$^3$ electric energy purely converted by heat. For the BTO based capacitors, the energy conversion performance does not degrade much over hundreds to thousands of cycles. Particularly, the Ce, Ca, Zr doped BTO capacitor generates 0.2 J/cm$^3$ electric energy in total of 1389 cycles, that is 3 orders of magnitudes bigger than the energy produced by other two materials. More importantly, the collected electric energy in external circuit almost equals to the electrostatic work done by the pyroelectric energy conversion. With this value, the pyroelectric device can be actually useful in real energy applications. 

\begin{table}[ht]
\caption{Performance of pyroelectric energy converters at their effective life. }
\begin{ruledtabular}
    \centering
 {\small   \begin{tabular}{lcccc}
            Material& $N^\ast$ & $W(N^\ast)$ J/cm$^3$ & $W_\pi(N^\ast)$ J/cm$^3$ & $W(N^\ast)/W_\pi(N^\ast)$ (\%)\\ \hline
         BTO & 502 & $6.24\times 10^{-4}$ & $8.36 \times 10^{-4}$ & 75\%\\
         BZT &  9 & $8.92 \times 10^{-4}$ & $0.0049$ & 18\%\\
tri-doped & 1389 & 0.1997 & 0.2018 & 99\%\\ 
    \end{tabular}
    \label{tab:work}
    }
\end{ruledtabular}
\end{table}

\section{Conclusion}

This paper underlies the influence of leakage on the performance of pyroelectric energy conversion using first-order phase-transforming ferroelectrics. Through experiments and numerical simulations done for BTO, BZT and tri-dope BTO, we show that the tri-doped BTO is most suitable for pyroelectric energy conversion for its singular polarization property and the much reduced leakage. Our model confirms that the degradation of energy conversion performance strongly depends on the leakage current density of the pyroelectric materials. To design a power-source-free pyroelectric energy conversion system that can continuously harvest electricity from the waste heat, both pyroelectric property and leakage current density should be taken into account. Our work provides useful reference in future development of high performance ferroelectric materials and the design of energy conversion systems.

\begin{acknowledgments}
C. Z. and X. C. thank the financial support of the HK Research Grants Council under Grant 16201118, 16201019 and the Bridge Gap Fund BGF.009.19/20 from HKUST Technology Transfer Center. 
\end{acknowledgments}

\bibliography{leakage_paper}

\end{document}